\documentclass[onecolumn,preprintnumbers,prb,fleqn]{revtex4}
\usepackage{graphicx,color}
\usepackage{amsmath,amssymb}
\newcommand{\be}{\begin{equation}} 
\newcommand{\ee}{\end{equation}} 
\newcommand{\bea}{\begin{eqnarray}}
\newcommand{\eea}{\end{eqnarray}}

\begin{document}
\title{Some Factors Leading to Asymmetry in Electronic Spectrum of Bilayer Graphene}
\author{Rupali Kundu}
\email{rupali@iopb.res.in}
\affiliation{ Institute of Physics, Bhubaneswar 751005 India.\\}
\date{\today}
\begin{abstract}
We have investigated the effects of inplane and interplane nearest neighbour overlap integrals ($s_0$ and $s_1$) and site energy difference between atoms in two different sublattices in the same graphene layer ($\Delta$) on the electronic dispersion of bilayer graphene within tight binding model. Also, modifications in bilayer graphene bands due to inplane next nearest neighbour interactions ($\gamma_{1i}$, $s_{1i}$) and next to next nearest neighbour interactions ($\gamma_{2i}$, $s_{2i}$) have been studied. It is observed that $s_1$ introduces further asymmetry in energy positions of top conduction band and bottom valence band at the $K$ point on top of the asymmetry due to $\Delta$. Moreover, $s_0$, $s_1$ as well as the other inplane coupling parameters induce noticable electron-hole asymmetry in the slope of the bands and changes in band widths. The density of states of bilayer graphene has also been calculated within the same model.
\end{abstract}
\maketitle
\section{Introduction}
Bilayer graphene is a system of two stacked hexagonal graphene sheets. Among all the carbon based materials of recent interest, bilayer graphene is of very much importance because this is the only two dimensional material in which the band gap between valence and conduction bands can be controlled by applying an external electric field perpendicular to the layers\cite{1,2,3,4,5,6} or by chemical doping of one of the layers\cite{7}. This makes it a potential candidate for future application in nanoelectronics. There have been an intensive research work going on over the last several years to understand the intresting behaviours it manifests. Since electronic band structure is one of the key ingredients to shed light on the understanding of the properties of a material, in particular, as far as electronic properties of the material are concerned, a significant amount of effort has been put to study the band structure of bilayer graphene\cite{6,7,8,9,10}. Regarding this the major concern was to see how the linear dispersion near the $K$ ponit in single layer graphene is modified in presence of interlayer coupling in bilayer graphene. It is now well established that even the slightest presence of the interlayer coupling kills the linearity of the dispersion of monolayer graphene and converts it to a parabolic one for the bilayer graphene. Moreover, most of the studies have talked about the symmetric nature of the valence and conduction bands of bilayer graphene. Only some very recent works\cite{10,11,12} have reported about asymmetry between the valence and conduction bands in this system. In this work we have investigated for some other parameters which also contribute appreciably towards asymmetry in bands of bilayer graphene. While the existing literature mostly discussed about interlayer coupling energies in bands of bilayer graphene and in graphite\cite{13}, we here report about asymmetry in band structure of bilayer graphene due to the inclusion of inplane and interplane nearest neighbour overlap integrals ($s_0$ and $s_1$) in the tight binding band structure calculation because presence of nearest neighbour overlap integral ($s_0$) gives quite a bit asymmetry in the band structure of monolayer graphene\cite{14,15}. We have also studied the effect of inplane next nearest neighbour (nnn) and next to next nearest neighbour (nnnn) hopping energies ($\gamma_{1i}$ and $\gamma_{2i}$) along with the corresponding overlap integrals ($s_{1i}$ and $s_{2i}$) on the band structure of this system in line with a similar work on single layer graphene\cite{16}. To calculate the bands we have used $\Delta=18$ meV and nearest neighbour interplane coupling energy $\gamma_1=400$ meV\cite{11}. For all the inplane parameters, we have used a set of parameters which we have determined by comparing the tight binding bands of single layer graphene with first principle results by considering the fact that magnitudes of the parameters should decrease with increasing distance in our earlier work\cite{17}. The article has been arranged in the following way: in section I we talk about the structure of bilayer graphene; section II focusses entirely on electronic structure of bilayer graphene with nearest neighbour inplane and interplane couplings, inplane next nearest and next to next nearest neighbour couplings; section III shows the density of states of bilayer graphene for the above mentioned bands and finally, summary and conclusions over the previous sections are given in section IV.
\subsection{Structure}
Bilayer graphene is a coupled system of two monolayers of graphene with usually an AB-stacking fashion of the layers, i. e. the atoms in one layer are not just on top of the corresponding atoms in the other layer, rather the arrangement is such that if one layer is projected on the other, the A-type (say) atoms coincide with the A-type atoms but the B-type atoms come at the center of the hexagons of the other layer\cite{8}. So A and B type atoms should not be treated as chemically equivalent carbon atoms because they belong to two different chemical environments. But since this site energy difference is very small ($\sim 18$ meV), they are frequently treated as chemically equivalent carbon atoms. The structure of AB stacked bilayer graphene and its brillouin zone are shown in figure (1). The distance between two atoms in a layer is 1.42\AA ~and the interlayer separation is 3.35\AA. The brillouin zone of bilayer graphene is same as that of single layer graphene because it does not have periodicity along z-direction. Bilayer graphene unit cell contains four atoms. The relevant hopping parameters, i. e. nearest neighbour inplane coupling energy ($\gamma_0$), nearest neighbour interplane coupling energy ($\gamma_1$), next nearest neighbour inplane coupling energy ($\gamma_{1i}$) and next to next nearest neighbour inplane coupling energy ($\gamma_{2i}$) have also been shown in the figure.
{\begin{figure}[h]
       \centering
       \includegraphics[width=10cm]{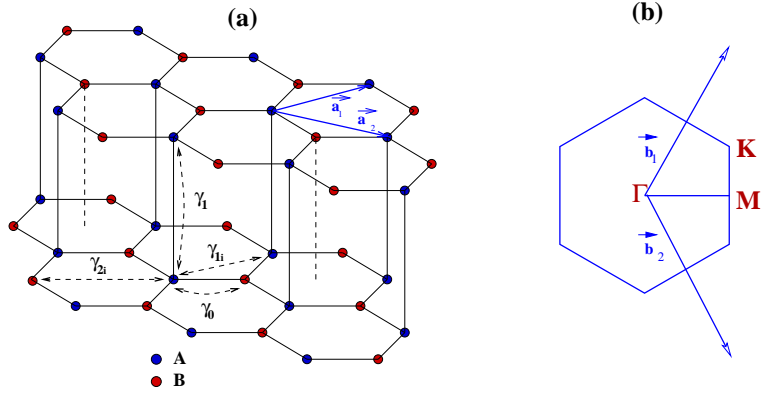}
        	\caption{(a) Structure of bilayer graphene with unit vectors $\vec a_1$, $\vec a_2$; the intralayer nn coupling energy ($\gamma_0$), interlayer nn coupling energy ($\gamma_1$), intralayer nnn coupling energy ($\gamma_{1i}$) and nnnn coupling energy ($\gamma_{2i}$). (b) Brillouin zone of bilayer graphene with unit vectors $\vec b_1$, $\vec b_2$ and the high symmetry directions.}
	\end{figure}}
\section{Electronic dispersion of Bilayer graphene}
In the following first we will be talking about some general aspects of the formalism of tight binding band structure of bilayer graphene, then the details of tight binding band dispersion under different situations like: ($i$) the simplest case with nearest neighbour inplane and interplane hopping of $\pi$ electrons; ($ii$) modifications in the bands due to the inclusion of overlap integrals coming from the same neighbours and due to site energy difference between A and B atoms in two different sublattices in the same layer; ($ii$) effect of inplane second nearest neighbour hopping and corresponding overlap integral; ($iv$) effect of inplane third nearest neighbour hopping and overlap integral on the band near the brillouin zone corner ($K$) point and over the whole brillouin zone, and the density of states of bilayer graphene corresponding to the above mentioned dispersions. Since bilayer graphene unit cell contains four atoms coming from two different sublattices of two layers, the Bloch wave functions for A and B type atoms for each layer\cite{9} are 
\bea
\Psi^{A_{i}}_{k}(r)&=&1/\sqrt{N}\sum_{A_{i=1,2}}\displaylimits \Phi_{A}(r-r_{A_{i}})e^{ik.r_{A_{i}}} \nonumber \\
\Psi^{B_{i}}_{k}(r)&=&1/\sqrt{N}\sum_{B_{i=1,2}}\displaylimits\Phi_{B}(r-r_{B_{i}})e^{ik.r_{B_{i}}}, \nonumber 
\eea
where $i$ refers to layer index, $\Phi^{'s}$ are $p_z$ atomic orbitals, N is the number of unit cells in the crystal, $C_{A_i}$ and $C_{B_i}$ are contributions to the total wavefunction coming from $i^{th}$ atom of A and B sublattices respectively. The total wave function is 
\be
\Psi_{k}(r)=\sum_{i=1, 2}\displaylimits C_{A_{i}}\Psi^{A_{i}}_{k}(r)+\sum_{i=1, 2}\displaylimits C_{B_{i}}\Psi^{B_{i}}_{k}(r). \nonumber
\ee
If H is the Hamiltonian, the general secular equation including overlap integrals becomes
\be
\left|\begin{array}{cccc}
H_{A_{1}A_{1}}-E(k)S_{A_{1}A_{1}} & H_{A_{1}B_{1}}-E(k)S_{A_{1}B_{1}} & H_{A_{1}A_{2}}-E(k)S_{A_{1}A_{2}} & H_{A_{1}B_{2}}-E(k)S_{A_{1}B_{2}}  \\
                H_{B_{1}A_{1}}-E(k)S_{B_{1}A_{1}} & H_{B_{1}B_{1}}-E(k)S_{B_{1}B_{1}} & H_{B_{1}A_{2}}-E(k)S_{B_{1}A_{2}} & H_{B_{1}B_{2}}-E(k)S_{B_{1}B_{2}}  \\
                H_{A_{2}A_{1}}-E(k)S_{A_{2}A_{1}} & H_{A_{2}B_{1}}-E(k)S_{A_{2}B_{1}} & H_{A_{2}A_{2}}-E(k)S_{A_{2}A_{2}} & H_{A_{2}B_{2}}-E(k)S_{A_{2}B_{2}}  \\
                H_{B_{2}A_{1}}-E(k)S_{B_{2}A_{1}} & H_{B_{2}B_{1}}-E(k)S_{B_{2}B_{1}} & H_{B_{2}A_{2}}-E(k)S_{B_{2}A_{2}} & H_{B_{2}B_{2}}-E(k)S_{B_{2}B_{2}}  \\
\end{array}\right|=0,
\ee
where $H^{'s}_{x_{i}y_{j}}$ and $S^{'s}_{x_{i}y_{j}}$ are defined as
\bea
H_{x_{i}y_{j}}&=&\left\langle \Psi^{x_{i}}_{k}|H|\Psi^{y_{j}}_{k}\right\rangle \text{ and} \nonumber \\
S_{x_{i}y_{j}}&=&\left\langle \Psi^{x_{i}}_{k}|\Psi^{y_{j}}_{k}\right\rangle \text{ respectively.} \nonumber
\eea
Here $x$, $y$ stand for both A and B; $i$, $j$ can take either of the values 1 and 2, and $S^{'s}$ are the overlap integrals.
\subsection{Modifications in the bands due to the inclusion of overlap integrals coming from inplane and interplane nearest neighbours and due to breaking of sublattice symmetry}
In this part, the band structure of bilayer graphene has been calculated in presence of nearest neighbour inplane and interplane coupling energies ($\gamma_0$, $\gamma_1$) along with the corrections coming from the corresponding overlap integrals ($s_0$, $s_1$). The sublattice energy difference ($\Delta$) has also been included. When solved, the eigenvalue equation readily gives the following solutions in case of the simplest possible situation with $\Delta$, $s_0$ and $s_1$ to be zero.
\bea
E_1(k)&=&-1/2\left[ \gamma_1 - \sqrt{\gamma^{2}_1+4\gamma^{2}_{0}g(k)}\right], \nonumber\\
E_2(k)&=&1/2\left[ \gamma_1 + \sqrt{\gamma^{2}_1+4\gamma^{2}_{0}g(k)}\right], \nonumber\\
E_3(k)&=&-1/2\left[ \gamma_1 + \sqrt{\gamma^{2}_1+4\gamma^{2}_{0}g(k)}\right] \text{ and} \nonumber\\
E_4(k)&=&1/2\left[ \gamma_1 - \sqrt{\gamma^{2}_1+4\gamma^{2}_{0}g(k)}\right], \nonumber\\
\text{where } g(k)&=&\left( 1+4\cos^2{(k_ya/2)}+4\cos{(\sqrt{3}k_xa/2)}\cos{(k_ya/2)}\right). \nonumber
\eea
Expansion of the dispersions around the $K\left( 2\pi/\sqrt{3}a, 2\pi/3a\right)$ point give
\bea
E_1(k)&=&-\gamma_1 - (3/4)\left(  \gamma^{2}_0/\gamma_1\right)  a^{2}(\triangle k)^{2}, \nonumber\\
E_2(k)&=& - (3/4)\left( \gamma^{2}_0/\gamma_1\right) a^{2}(\triangle k)^{2}, \nonumber\\
E_3(k)&=& + (3/4)\left( \gamma^{2}_0/\gamma_1\right) a^{2}(\triangle k)^{2} \text{ and} \nonumber\\
E_4(k)&=&\gamma_1 + (3/4)\left( \gamma^{2}_0/\gamma_1\right) a^{2}(\triangle k)^{2}, \nonumber
\eea
where $\triangle k$ is a small change in $k$ around the $K$ point.
From the above results we see that there are four $\pi$ bands in bilayer graphene. All the bands are parabolic near the brillouin zone corner. Two of the bands are degenerate with zero energy at the $K$ point. -$\gamma_1$ and $\gamma_1$ are the energies of the other two bands at that point, i. e. $\pm \gamma_1$ is the energy separation between the highest conduction band and the lowest valence band at that point. Also the valence and conduction bands are symmetric over the entire brillouin zone. Since each atomic site has one $p_z$ electron, the fermi energy ($E_F$) for undoped bilayer graphene is at zero energy. The bands are plotted in figure (2) along with the bands with nearest neighbour inplane overlap integral ($s_0$), nearest neighbour interplane overlap integral ($s_1$) and sublattice asymmetric energy ($\Delta$).
Including finite values of $\Delta$, $s_0$ and $s_1$ the secular equation becomes
\be
\left|\begin{array}{cccc}
 E_0+\Delta-E(k)& a-b E(k)& \gamma_1-s_1 E(k)& 0\\
  c-d E(k)& E_0-E(k)& 0 & 0\\
 \gamma_1-s_1 E(k)& 0 & E_0+\Delta-E(k)& c-d E(k)\\
  0 & 0 & a- b E(k)& E_0-E(k)\\
\end{array}\right|=0,
\ee
where $a=\gamma_0 f(k)$, $b=s_0 f(k)$, $c=\gamma_0 f^*(k)$ and $d=s_0 f^*(k)$. Following are the eigen solutions to the above equation.
\bea
E_{1,2}(k)&=& \left[ -B_1\pm \sqrt{B^{2}_1-4A_1C_1}\right] /2A_1, \\
\text{where } A_1&=& 1-bd+s_1, ~~B_1= ad+bc-\gamma_1-s_1 E_0-2 E_0 -\Delta, ~~ C_1 = \gamma_1 E_0 - ac + E_0(E_0 +\Delta)\nonumber\\
\text{and }E_{3,4}(k)&=& \left[ -B_2\pm \sqrt{B^{2}_2-4A_2C_2}\right] /2A_2, \\
\text{where } A_2&=& 1-bd-s_1, ~~ B_2 = ad+bc+\gamma_1+s_1 E_0-2 E_0 - \Delta, ~~ C_2 = -\gamma_1 E_0 - ac + E_0(E_0 + \Delta).\nonumber
\eea
{\begin{figure}[h]
       \centering
       \includegraphics[width=12cm]{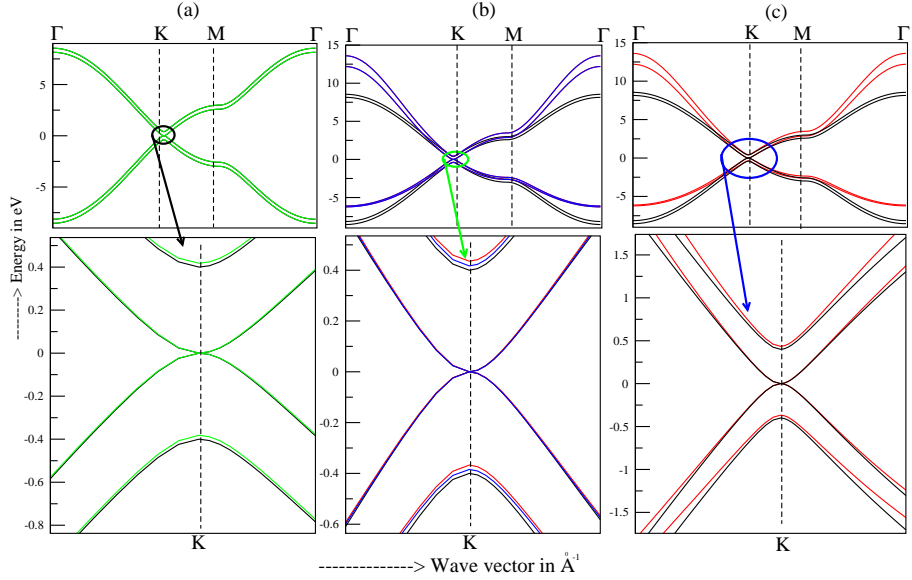}
        	\caption{Electronic spectra of bilayer graphene. For all three cases, in the upper panel we have plotted the bands along all the high symmetry directions of the brillouin zone and the lower panels show zoomed versions of the corresponding upper panels near the $K$ point. In all the panels the black curves represent the symmetric bands, i. e. bands in presence of nearest neighbour inplane ($\gamma_0$) and interplane ($\gamma_1$) hopping energies. In (a) the green curves contain informations due to the presence of sublattice asymmetry ($\Delta$), clearly visible from the lower panel; in (b) the blue curves have the effect of nearest neighbour inplane ($s_0$) and interplane ($s_1$) overlap integrals but zero $\Delta$, whereas the red curves are plotted with non-zero values of $s_0$, $s_1$ and $\Delta$. Effects of these parameters at the $K$ point are visible from the lower panel. In (c) the bands in presence of finite values of $s_0$, $s_1$ and $\Delta$ (red curve) are compared with the symmetric bands (black curve) over the whole brillouin zone (upper panel) and nearer to the $K$ point within optical energy range (lower panel).}
	\end{figure}}
We first discuss the effect of the asymmetry term ($\Delta$) on the spectra at $K$ point. The two bands which are having degeneracy at $K$ point are not affected by $\Delta$ at that point but the other two bands appear at energies $\Delta-\gamma_1$ and $\Delta+\gamma_1$, i. e. though the separation is still $\pm \gamma_1$, they become asymmetric regarding energy positions at that point. This term has negligible effect on the slope of the curves. The lower panel of (a) in fig. (2) shows that the lower valence band comes closer to $E_F$ whereas the upper conduction band moves away from $E_F$ when the sublattice asymmetry is taken care of. Next we illustrate the effect of overlap integrals $s_0$ and $s_1$ at $K$ point. It is observed that the inplane nearest neighbour overlap integral $s_0$ has no effect at $K$ point because it always appear in product with $g(k)$ which vanishes at $K$. But the interlayer overlap integral $s_1$, though not altering the degenerate bands, affects those bands lying further from $E_F$ quite significantly. The upper lying band shifts further to $-\gamma_1/(1-s_1)$ and the lower lying band comes closer to $\gamma_1/(1+s_1)$ (blue curve in lower panel of (b) in fig. (2)). The energy separation between them becomes $2\gamma_1/(1-s^{2}_{1})$. When $\Delta$ and overlap integrals are considered together, the upper lying band shifts even further to $(\Delta -\gamma_1)/(1-s_1)$ and the lower lying band comes more close to $(\Delta + \gamma_1)/(1+s_1)$ (red curve in lower panel of (b) in fig. (2)). Under this condition the above energy separation becomes $2(\gamma_1-s_1 \Delta)/(1-s^{2}_{1})$. Moreover, $s_0$ and $s_1$ play important role in changing the slope of the curves. They push off the conduction bands and pull in the valence bands almost over the entire energy range, particularly near the $\Gamma$ point the effect is most prominent (red curves in upper panel of (c) in fig. (2)). With a close look over a smaller energy range (lower panel of (c)) we see that the conduction bands are repelling each other towards higher energy side whereas the valence bands have the tendency to merge.
\subsection{Effect of inplane second nearest neighbour hopping and corresponding overlap integral on the spectrum of bilayer graphene}
In this part we have studied the changes in the above dispersions by taking care of inplane second nearest neighbour transfer integral ($\gamma_{1i}$) and the corresponding overlap integral ($s_{1i}$). Since the inplane next nearest neighbour atoms belong to same sublattice in a plane, only the entities $H_{A_{1}A_{1}}$, $S_{A_{1}A_{1}}$; $H_{A_{2}A_{2}}$, $S_{A_{2}A_{2}}$; $H_{B_{1}B_{1}}$, $S_{B_{1}B_{1}}$ and $H_{B_{2}B_{2}}$, $S_{B_{2}B_{2}}$ get modified and changes occur in $A_1$, $B_1$, $C_1$ and $A_2$, $B_2$, $C_2$ as follows
\bea
A_1&=& (1 + s_{1i} u(k))^2 - bd  +s_1(1 + s_{1i} u(k)), \nonumber\\
B_1&=& ad+bc - \gamma_1(1 + s_{1i} u(k)) - s_1 (E_0+\gamma_{1i}u(k))-(1 + s_{1i} u(k))(2 E_0 + 2\gamma_{1i}u(k) + \Delta), \nonumber\\
C_1&=& \gamma_1 (E_0 + \gamma_{1i}u(k)) - ac + (E_0 + \gamma_{1i}u(k))(E_0 +\gamma_{1i}u(k)+\Delta), \text{ and} \nonumber\\
A_2&=& (1 + s_{1i} u(k))^2 - bd - s_1(1 + s_{1i} u(k)), \nonumber\\
B_2&=& ad+bc + \gamma_1(1 + s_{1i} u(k)) + s_1 (E_0+\gamma_{1i}u(k))-(1 + s_{1i} u(k))(2 E_0 + 2\gamma_{1i}u(k) + \Delta), \nonumber\\
C_2&=& -\gamma_1 (E_0 + \gamma_{1i}u(k)) - ac + (E_0 + \gamma_{1i}u(k))(E_0 +\gamma_{1i}u(k)+\Delta),\nonumber \\
\text{where }u(k)&=&2\cos{(k_ya)}+4\cos{(k_xa\sqrt{3})}\cos{(k_ya/2)}. \nonumber
\eea
\subsection{Modification due to inplane third nearest neighbour hopping energy and overlap integral on the bands of bilayer graphene}
Inplane third nearest neighbour atoms belonging to the other sublattice incorporate changes to the quantities $H_{A_{1}B_{1}}$, $S_{A_{1}B_{1}}$; $H_{A_{2}B_{2}}$, $S_{A_{2}B_{2}}$ and their complex conjugates which inturn modify $A_1$, $B_1$, $C_1$ and $A_2$, $B_2$, $C_2$ through changes in $a$, $b$, $c$ and $d$. Third nearest neighbour inplane transfer energy and overlap integrals are denoted as $\gamma_{2i}$ and $s_{2i}$. Modified $a$, $b$, $c$ and $d$ appear as
\be
a=\gamma_0 f(k)+\gamma_{2i} f_1(k), ~~b=s_0 f(k)+s_{2i}f_1(k), ~~ c=\gamma_0 f^*(k)+\gamma_{2i} f^*_1(k)~~ \text{and}~~d=s_0 f^*(k)+s_{2i}f^*_1(k),\nonumber
\ee
where $f_1(k)=e^{ik_xa/\sqrt{3}}2\cos{k_ya}+e^{-2ik_xa/\sqrt{3}}$.\\
So the eigen solutions in this case are same as equation (3) and (4) with a different set of $A_1$, $B_1$, $C_1$ and $A_2$, $B_2$, $C_2$ due to differences appearing in $a$, $b$, $c$ and $d$. 
To illustrate the effects of inplane next nearest neighbours and next to next nearest neighbours on the spectra of bilayer graphene, all the bands are plotted together with the symmetric spectra (black curves) and the spectra with corrections due to nearest neighbour inplane and interplane overlaps (red curve) in figure (3).
\vspace{0.5 cm}
\begin{figure}[h]
      \centering
       \includegraphics[width=10cm]{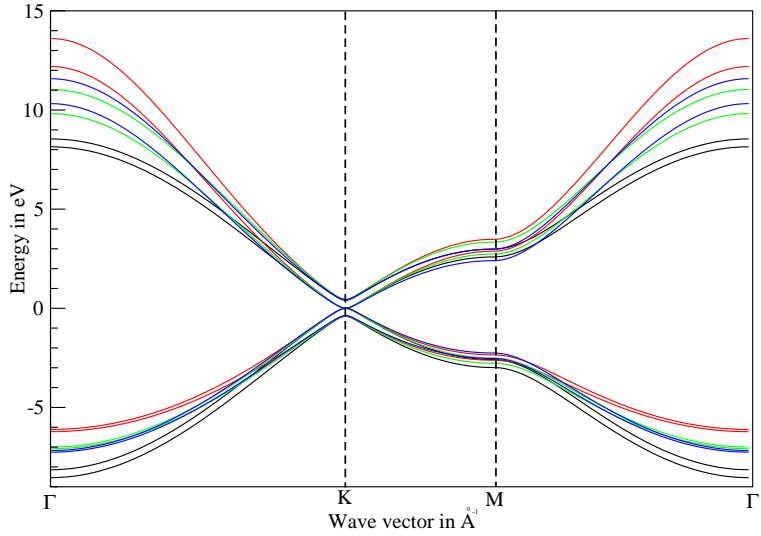}
        	\caption{Electronic dispersion bilayer graphene in presence of nearest neighbour inplane and interplane transfer integrals (black curve), nearest neighbour inplane and interplane transfer integrals, overlap integrals and sublattice asymmetric energy (red curve), inplane next nearest neighbour interactions (green curve) and inplane next to next nearest neighbour interactions (blue curve). The parameters used for these bands are given in Table I.}
	\end{figure}
From the plots shown in fig. (3) and from plots not shown here it is very clear that presence of any of the parameters ($s_0$, $s_1$, $\gamma_{1i}$, $s_{1i}$, $\gamma_{2i}$, $s_{2i}$) or all of them together introduces asymmetry of different amounts on the symmetric bands due to $\gamma_0$ and $\gamma_1$. While $\Delta$ makes the upper conduction and lower valence band positions asymmetric at $K$ point, others give asymmetry to the bands almost over the whole brillouin zone. We have already discussed the effects of $s_0$, $s_1$ on the spectra with $\gamma_0$ and $\gamma_1$ in the previous section. Now, we describe the effects due to inplane next nearest neighbours ($\gamma_{1i}, s_{1i}$) on top of $s_0$, $s_1$. There is no significant change in energy positions of the top conduction band and the bottom valence band at the $K$ ponit in presence of inplane nnn and nnnn interactions. For the valence bands the nnn bands are nearer to the symmetric bands within $\sim$ 2 eV but the nnnn bands are closer to the modified bands due to $s_0$ and $s_1$. For the conduction bands the scenario is opposite within the same energy range. Beyond that the nnn bands move away from the symmetric bands, though in opposite directions for valence and conduction bands with respect to fermi energy, nnnn bands start moving towards the symmetric bands such that the nnn and nnnn valence bands almost coincide and the co duction bands keep a small separation among themselves around $\Gamma$ point. The values of the parameters used for the plotted bands are given in Table I.
\begin{table}[h]
\caption{Tight binding parameters}
\begin{center}
\begin{tabular}{|c| r|r|r|r|r|r|r|r|r|r|}
\hline\hline
Curves & $E_0$(eV) & $\Delta$(eV) & $\gamma_0$(eV) & $\gamma_1$(eV) & $\gamma_{1i}$(eV) & $\gamma_{2i}$(eV) & $s_0$ & $s_1$ & $s_{1i}$ & $s_{2i}$  \\ 
\hline
Black &  &  &-2.78 & -0.4 & &  &  & &  &   \\ 
\hline
Red &  & 0.018 &-2.78 & -0.4 & &  & 0.117 &0.04 & &   \\   
\hline
Green & -0.45 & 0.018 &-2.78 & -0.4 &-0.15 &  & 0.117 &0.04 & 0.004 &  \\ 
\hline
Blue & -0.45 & 0.018 &-2.78 & -0.4 &-0.15 & -0.095 & 0.117 &0.04 & 0.004 & 0.002 \\ 
\hline
\end{tabular}
\end{center}
\end{table}
In presence of inplane next nearest neighbour interactions there are three choices of $E_0$ ($=3\gamma_{1i}, ~ 3\gamma_{1i}-\Delta \pm \gamma_1$) for which the fermi energy comes at zero. For other choices of $E_0$ the the fermi level will shift from zero. In presence of inplane next to next nearest neighbours the choices of $E_0$ for which fermi energy will be at zero are same as in case of nnn interactions.
\subsection{Density of states of bilayer graphene}
The density of states (DOS) of bilayer graphene corresponding to the bands in fig. (3) with the parameters in Table I are shown in figure (4).
{\begin{figure}[h]
       \centering
       \includegraphics[width=7cm]{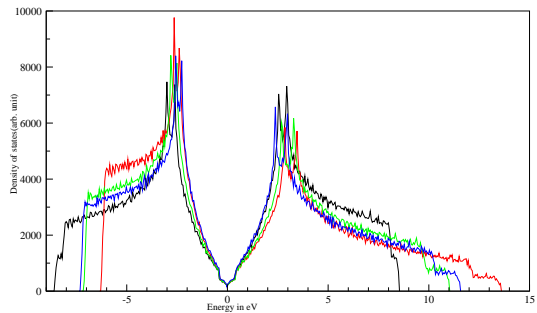}
        	\caption{Density of states of bilayer graphene in presence of nearest neighbour inplane and interplane transfer integrals (black curve), nearest neighbour inplane and interplane transfer integrals, overlap integrals and sublattice asymmetric energy (red curve), inplane next nearest neighbour interactions (green curve) and inplane next to next nearest neighbours (blue curve). The parameters used in these curves are given in Table I.}
	\end{figure}}
From the curves it is evident that the finite, though small, density of states (DOS) at the fermi energy of bilayer graphene remain unaffected due the presence of factors leading to asymmetry in bands. Further, the slope of the density of states curves within $< 0.5$ eV does not change due to the above mentioned factors. Rather, they have prominent effect in bringing asymmetry in band widths. To illustrate, we see that when only nearest neighbour inplane and interplane transfer energies ($\gamma_0$ and $\gamma_1$) are there, valence band DOS and conduction band DOS (black curve) are exactly symmetric with respect to the fermi energy at zero, i.e. band widths are same. Also the van-Hov singularities are at symmetric positions, the scenario being consistent with the corresponding band structure. As soon as the nearest neighbour inplane and interplane overlap integrals ($s_0$ and $s_1$) are taken into account the valence band (VB) DOS and conduction band (CB) DOS start becoming asymmetric, i.e. the band widths become different: valence band becomes narrow and conduction band widens. Moreover, the van-Hov singularities in VB come closer to $E_F$ whereas those in CB move away from $E_F$. In presence of inplane second nearest neighbours CB becomes slightly narrow and VB slightly wide compared to the previous case, though VB is still narrower compared to CB. There is no significant change in the positions of the van-Hov singularities in this case when compared with the previous one. Inplane third nearest neighbours do not have much effect on the widths of the bands on top of that of the inplane second nearest neighbours but bring the van-Hov singularities slightly nearer to the fermi energy. It is also observed that the sublattice asymmetric energy ($\Delta$) does not have any significant effect on the density of states of bilayer graphene.
\section{Summary and Conclusions}
To summarize, we have illustrated the effects of various parameters governing electron-hole asymmetry in the band structure of bilayer graphene within tight binding model. When compared the role of site energy difference ($\Delta$) between A and B sublattices in the same graphene layer on the electronic spectra of single layer graphene and bilayer graphene, a distinct difference is observed between the two systems. Sublattice asymmetry in monolayer graphene introduces a gap in the spectra at the $K$ point whereas in bilayer it does not induce gap in the spectra, rather it gives an asymmetry in the energy positions of the top valence and bottom conduction bands with respect to the energy at which the other two bands are degenerate. Moreover, in presence of $\Delta$ the gap between top valence and bottom conduction bands at the $K$ point remain intact to $\pm \gamma_1$ which is the separation even without $\Delta$. Apart from $\Delta$ the other important factor which contributes significantly to the positions of the top valence and bottom conduction bands at $K$ point is the interlayer nearest neighbour overlap integral ($s_1$). Hence, we find that $E_c$(top) and $E_v$(bottom) are functions of $\Delta$ and $s_1$ both at $K$ point. Regarding this ref. (11) discusses only about the dependence on $\Delta$. Our study suggests for the consideration of a model containing both $\Delta$ and $s_1$ for a more accurate determination of $\Delta$ from experimental results. Further, we observe a considerable change in the slope of the bands in presence of nearest neighbour inplane and interplane overlap integrals ($s_0$ and $s_1$) compared to those with nearest neighbour inplane and interlayer coupling energies ($\gamma_0$ and $\gamma_1$) only. Ref. (11) has discussed about the induction of electron-hole asymmetry in the slope of valence and conduction bands due to next nearest neighbour interlayer coupling energy ($\gamma_4$) but without overlap integrals. It is noted there that with finite $\gamma_4$ two conduction bands are closer and the valence bands are further apart at $k$ values away from the $K$ point compared to the bands without $\gamma_4$. We observe that finite values of $s_0$ and $s_1$ also give electron-hole asymmetry in the slope of valence and conduction bands but with an opposite trend compared to the bands with $\gamma_4$. With finite values of $s_0$ and $s_1$ the conduction bands move futher from each other and the valence bands come closer for momentum values away from the $K$ point value. It could be concluded from this comparative study that the combined effect of $\gamma_4$, $s_0$ and $s_1$ could be somewhat balancing between the above two cases, may not be a complete balance to get back the symmetric spectra with only $\gamma_0$ and $\gamma_1$ but the degree of electron-hole asymmetry in slope of the bands will get modified. A model including $\gamma_4$, $s_0$ and $s_1$, though very complicated to handle, may lead to more accurate determination of the important parameter like $\gamma_1$ when comparing experimental results having asymmetry in electron and hole sides. Now comes the inplane next nearest neighbour interactions ($\gamma_{1i}$ and $s_{1i}$). With $\gamma_{1i}$ and $s_{1i}$ the trend of electron-hole asymmetry in slope of bands remain similar as that with $s_0$ and $s_1$ but the asymmetry in valence and conduction band widths near $\Gamma$ point is reduced compared to that with zero values of $\gamma_{1i}$ and $s_{1i}$. Moreover, in this case if the site energy term $E_0$ is not chosen properly there is a shift in fermi energy\cite{12}. Inplane third nearest neighbour interactions ($\gamma_{2i}$ and $s_{2i}$) do not affect much on top of inplane second nearest neighbours except very little change in band withs at $\Gamma$ point and slight modifications in the slope of the bands. Hence, as far as electron-hole asymmetry in slope of valence and conduction bands are concerned, a Hamiltonian including $s_0$, $s_1$ and inplane second nearest neighbour interactions ($\gamma_{1i}$ and $s_{1i}$) could be sufficient to interpret experimental results (e. g. cyclotron resonance data) with asymmetry in electron and hole sides. Also, the inplane third nearest neighbour interactions could be more useful in determining all the above mentioned parameters by fitting bilayer graphene bands with first principle results or with angle resolved photoemission(ARPES) data over the whole brillouin zone.
\section{Acknowledgement}
Many important suggestions and discussions from Prof. S. G. Mishra and also useful discussions and cooperation for this work from Prof. B. R. Sekhar are greatly acknowledged.

\end{document}